\documentclass[runningheads,a4paper]{llncs}

\usepackage{microtype}
\usepackage{amssymb}
\usepackage{bbm}
\usepackage{nicefrac}
\usepackage{units}
\usepackage{graphicx}
\usepackage{url}
\usepackage{booktabs}
\usepackage[binary-units]{siunitx}
\usepackage{fancyvrb}
\usepackage[textsize=footnotesize]{todonotes}
\usepackage{hyperref}
\usepackage{tikz}
\usepackage{xfrac}
\usepackage{relsize}
\usepackage{xcolor-solarized}

\usetikzlibrary{calc, positioning, fit, backgrounds}

\usepackage{pdftexcmds}
\makeatletter
\ifcase\pdf@shellescape
  \or
  \usepackage{minted}\or
  \fi
\makeatother

\urldef{\mailsa}\path|{a.herten,t.hater,w.klijn,d.pleiter}@fz-juelich.de|
\newcommand{\keywords}[1]{\par\addvspace\baselineskip
\noindent\keywordname\enspace\ignorespaces#1}

\title{Performance Comparison for Neuroscience Application Benchmarks}

\titlerunning{Performance Comparison for Neuroscience Application Benchmarks}

\author{%
  Andreas Herten$^1$ \and 
  Thorsten Hater$^1$ \and 
  Wouter Klijn$^1$ \and 
  Dirk Pleiter$^1$
}
\authorrunning{A. Herten et al.}

\institute{%
  $^1$Forschungszentrum J\"ulich, JSC, 52425 J\"ulich, Germany \\
  \mailsa
}

\toctitle{Lecture Notes in Computer Science}
\tocauthor{%
  Andreas Herten,
  Thorsten Hater,
  Wouter Klijn and
  Dirk Pleiter
}

\DeclareSIUnit\flop{Flop}
\DeclareSIUnit\cycle{cycle}

\begin{document}

\mainmatter

\maketitle

\begin{abstract}
Researchers within the Human Brain Project and related projects have in the last couple of years
expanded their needs for high-performance computing infrastructures.
The needs arise from a diverse set of science challenges that range from large-scale simulations
of brain models to processing of extreme-scale experimental data sets.
The ICEI project, which is in the process of creating a distributed infrastructure optimised for
brain research, started to build-up a set of benchmarks that reflect the diversity of applications
in this field.
In this paper we analyse the performance of some selected benchmarks on an IBM POWER8 and
Intel Skylake based systems with and without GPUs.

\end{abstract}

\section{Introduction}

As new computational and data science communities emerge, needs arise for having a benchmark suite reflecting the requirements of the respective communities, also for potential procurement of IT equipment. At the same time, experience needs to be collected regarding  performance observations on different types of hardware architectures.
In this contribution we address the latter by selecting a recently developed benchmark suite
and comparing performance results obtained on servers based on different processor architectures,
namely POWER8 and Skylake, with and without GPU acceleration.

The science community, on which we focus here, is the brain research community organised in
the Human Brain Project (HBP).\footnote{\url{https://www.humanbrainproject.eu/en/}}
HBP is a large-scale flagship project funded by the European Commission working towards
the realisation of a cutting-edge research infrastructure that will allow researchers
to advance knowledge in the fields of neuroscience, computing, and brain-related medicine.
As part of HBP, the ICEI project (Interactive Computing e-infrastructure for the Human Brain
Project) was started in early 2018.
This project plans to deliver a set of e-infrastructure services that will be federated to
form the Fenix Infrastructure.\footnote{\url{https://fenix-ri.eu/}}
The European ICEI project is funded by the European
Commission and is formed by the leading European Supercomputing Centres BSC in Spain,
CEA in France, CINECA in Italy, CSCS in Switzerland and J\"{u}lich Supercomputing Centre (JSC) in Germany.
To guide the creation of this infrastructure, the ICEI project started to build-up the
``ICEI Application Benchmark Suite'', which we use for this contribution.

This paper is organised as follows:
We start with giving an overview of the ICEI benchmark suite as well as the systems used for
collecting performance results in section~\ref{sec:benchmark_suite} and \ref{sec:test_systems},
respectively.
The obtained results are documented in section~\ref{sec:results}.
We finally provide a summary and conclusions in
section~\ref{sec:summary}.

\section{ICEI benchmark suite}\label{sec:benchmark_suite}

The components of the ``ICEI Application Benchmark Suite'' have been chosen such that it represents
the breadth of research within HBP. The subset of benchmarks, which we consider in this paper,
is directly based on real-life applications.
\emph{NEST} \cite{Linssen2018} is one of several simulators that became part of the benchmark
suite. It is a simulator for spiking neural network models that focuses on the dynamics,
size, and structure of neural systems rather than on the exact morphology of individual neurons.
Recently, a significantly improved uptake of this simulators in different areas of brain research
has been observed. NEST is a community code with an active user base. A key design goal is
extreme (weak) scalability, which could be demonstrated different supercomputers
(see, e.g., \cite{10.3389/fninf.2018.00002}).
The program is written in C++ and Python, and uses MPI and OpenMP for parallelisation.

Unlike NEST, \emph{Arbor} \cite{DBLP:conf/pdp/AkarCKKKPY19} is a simulation library for networks
of morphologically detailed neurons.
Simulations progress by taking half time steps for updating the states of the cells. This allows overlapping the exchange of the spikes generated by the cells.
During the communication of spikes with other cells, similar operations need to be performed as in case of NEST.
The performance in this step will mainly depend on memory and network performance.
The step of updating the cells is, however, more compute intensive and can potentially benefit from
compute acceleration through SIMD pipelines or GPUs.
Cells are represented as trees of line segments, on which partial differential equations
for potentials are solved using the finite-volume method.
For complex cell models the second step will dominate application performance. Arbor is mainly written in C++ and employs MPI and OpenMP as well as CUDA for parallelisation.

The Virtual Brain (TVB) \cite{TVB,Jirsa2010TowardsTV,doi:10.1089/brain.2012.0120}
is an application that aims at full brain network simulation.
It uses mesoscopic models of neural dynamics, which model whole brain regions.
For the interconnection of the different regions structural connectivity data sets are used.
The application can generate outputs on different experimental modalities
(for instance EEG or fMRI) and thus allows to compare simulated and experimental data.
To enable exploitation of supercomputers, a new version of the application is
being implemented, which is called \emph{TVB-HPC}.
TVB-HPC is written in Python and aims to automatically produce code for different targets, 
including processor architectures with SIMD pipelines or GPU accelerators. One of the 
targets is Numba \cite{Lam:2015:NLP:2833157.2833162}, which is a tool that translates 
Python functions to optimized machine code. TVB-HPC is using Numba for the benchmark at hand to just-in-time-compile Python code to CPU assembly. MPI is used to distribute tasks.

While the previous three applications enable different kind of brain simulations,
the remaining applications, which have been used for the ``ICEI Application Benchmark Suite''
and are considered here, address data analysis tasks.

This includes \emph{ASSET} \cite{10.1371/journal.pcbi.1004939},
which is part of the Elephant (Electrophysiology Analysis Toolkit).
Elephant is a library comprising a set of tools for analysing spike train data and other
time series recordings obtained from experiments or simulations.
Elephant is written in Python and relies on NumPy and SciPy for numerical tasks
and MPI/mpi4py parallelisation.
The tool ASSET (Analysis of Sequences of Synchronous EvenTs)
was developed to automatise processing of spike data for
sequences of synchronous spike events. In the ASSET benchmark at hand, one of the main compute kernels is compiled with Cython.

Another type of data processing challenge occurs in the context of analysis
of high-resolution images of histological brain sections.
To automatise the analysis of such images, applications based on deep learning
techniques have been developed \cite{7950666}.
The \emph{Neuroimaging Deep Learning} benchmark is derived from one such application.
It is based on TensorFlow in combination with Horovod for parallelisation, using TensorFlow's GPU backend in the benchmark presented here.

\section{Test systems}\label{sec:test_systems}

The ``ICEI Application Benchmark Suite'' has been executed on a variety of systems
to improve portability and collect performance results for different architectures.
Here we focus on results obtained on two systems installed at J\"{u}lich Supercomputing Centre:
\begin{itemize}
\item JURON is a pilot system dedicated to users from HBP, which was delivered by IBM and NVIDIA
      in the context of a pre-commercial procurement that was executed during an the initial phase
      of the HBP.
\item JUWELS is a flagship cluster system at JSC, which is one of the PRACE Tier-0 systems that
      are accessible for European researchers at large.
\end{itemize}

The 18 compute nodes of JURON are IBM S822LC servers (also known under the codename \emph{Minsky}).
Each node comprises two IBM POWER8 processors and four NVIDIA P100 GPUs.
Each group of one processor and two GPUs is interconnected via NVLink links.
The compute nodes are connected via Mellanox ConnectX-4 Infiniband EDR network adapters
to a single switch.
In the following we use the term ``CPU-only nodes'' when referring to JURON nodes
where the GPUs are not used.

The JUWELS cluster comprises \num{2511} CPU-only and \num{48} GPU-accelerated compute nodes.
Each comprises two Intel processors of the Skylake generation.
The GPU-accelerated nodes are additionally equipped with four NVIDIA V100 GPUs.
While the four GPUs are interconnected via NVLink in an all-to-all topology,
each GPU is only connected via one PCIe Gen3 link to one of the CPUs.
The compute nodes furthermore comprise a single Mellanox ConnectX-5 Infiniband EDR
network adapter through which they are interconnected using a fat-tree topology.

A more detailed comparison of the hardware capabilities of the nodes used for either system
are collected in Table~\ref{tab:hwpar}.
As the benchmarks considered here are compute-only (any time spent in I/O is not considered),
we do not report on I/O capabilities of both systems.

\begin{table}[ht]
  \setlength{\tabcolsep}{7pt}
  \caption{Comparison of node-level aggregated hardware parameters.}
  \label{tab:hwpar}
  \begin{center}
    \begin{tabular}{lcc}
      \toprule
                                        & JURON         & JUWELS                        \\
      \midrule
      Type of CPU                       & POWER8        & Intel Xeon Platinum 8168 /    \\
                                        &               & Intel Xeon Gold 6148 (GPU-acc.) \\
      Number of CPUs                    & 2             & 2                             \\
      Number of cores                   & 20            & 48 / 40                       \\
      Number of hardware threads        & 160           & 96 / 80                       \\
      SIMD width / \si{\bit}                  & 128           & 512                           \\
      Throughput / \si{\flop\per\cycle}          & 160           & \num{1536} / \num{1280}                 \\
      Memory capacity / \si{\gibi\byte}   & 256           & $\geq$96                        \\
      Memory bandwidth / \si{\giga\byte\per\second} & 230           & 255                           \\
      LLC capacity / \si{\mebi\byte}      & 160           & 66 / 27.5                     \\
      \midrule
      Number of GPUs                    & 4             & -- / 4                        \\
      Type of GPU                       & P100 SXM2     & V100 SXM2                     \\
      Throughput / \si{\flop\per\cycle}          & \num{14336}        & \num{20480}                        \\
      Memory capacity / \si{\gibi\byte}   & 64            & 64                            \\
      Memory bandwidth / \si{\giga\byte\per\second} & \num{2880}         & \num{3600}                         \\
      \bottomrule
    \end{tabular}
  \end{center}
\end{table}

\section{Results}\label{sec:results}

In this section we document selected results for the benchmark derived from the applications
introduced in section~\ref{sec:benchmark_suite}, which have been obtained on the systems
introduced in section~\ref{sec:test_systems}.

\subsection{NEST}

The benchmark is based on Version 2.14 of 
NEST~\cite{nest214}.\footnote{\url{https://github.com/nest/nest-simulator.git}}
Simulations are performed using a randomly connected network of \num{112500} neurons with
each neuron being connected to about \SI{10}{\percent} of the other neurons.
While the problem size is kept fixed, the number of MPI tasks and OpenMP threads can be varied.
Internally, NEST defines virtual processes (VP) and assigns one VP to each thread.
The application first builds a network, i.e. creates all neurons and connects them.
In a second step simulations are performed.
Here, \SI{1000}{\milli\second} biological time are simulated.
For this paper the GCC C++ compiler version 5.4.0 and 5.5.0 have been used on JURON and JUWELS, 
respectively.
Since NEST does not support GPU acceleration, we use the CPU-only nodes on JUWELS.

\begin{figure}[ht]
    \begin{center}
        \includegraphics[width=\textwidth]{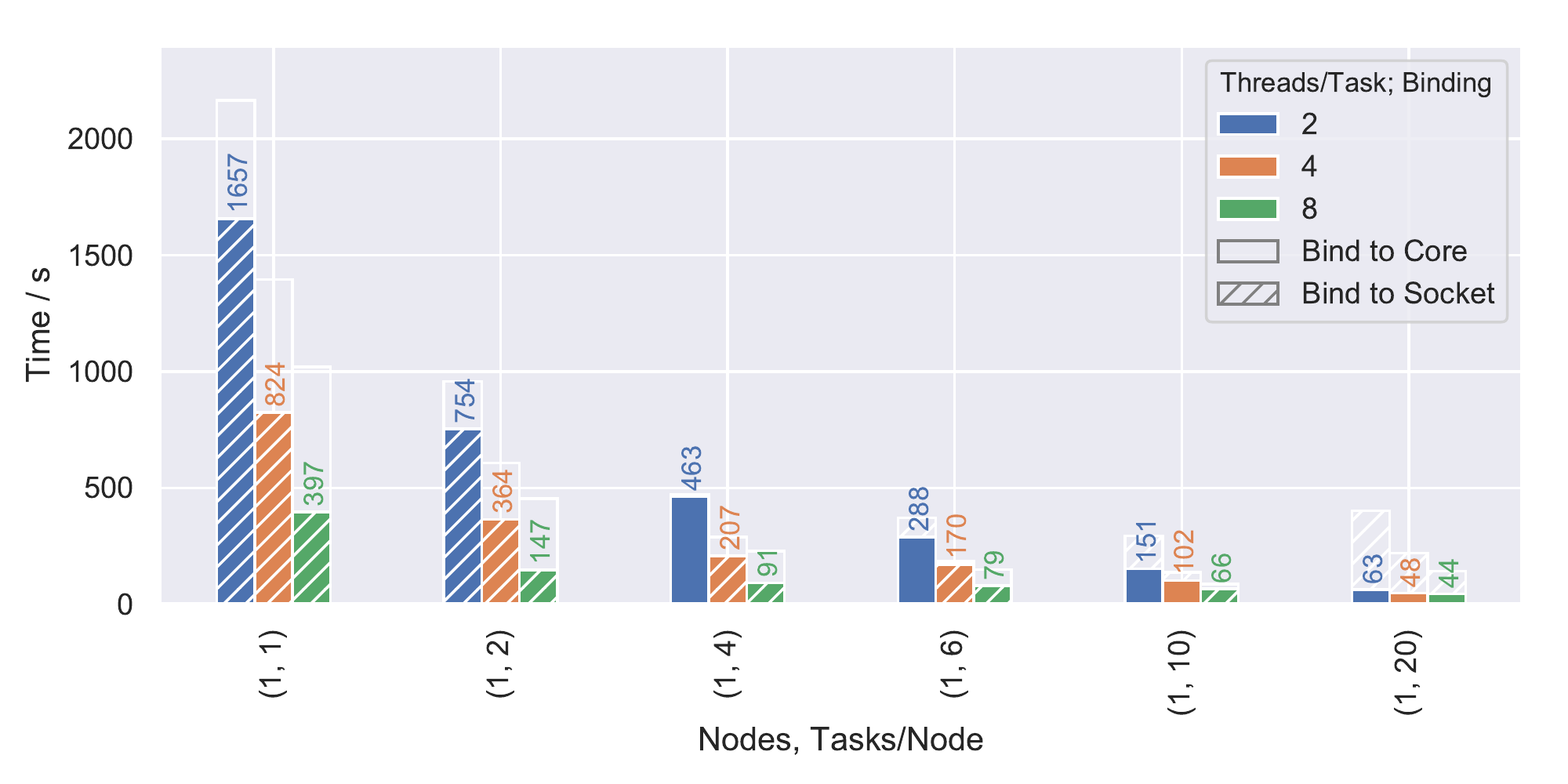}
    \end{center}
    \caption{\label{fig:nest:singlenode:juron}NEST benchmark -- JURON: Simulation time 
    on a single JURON node for different numbers of virtual processes (VP, the product of the number of nodes, tasks per node, and threads per task). Different strategies of mapping and binding MPI tasks to the system have been tested and the best performing configuration selected; a shaded bar denotes \emph{binding to socket}, a solid bar denotes \emph{binding to core}.}
\end{figure}

\begin{figure}[ht]
  \centering
  \includegraphics[width=\textwidth]{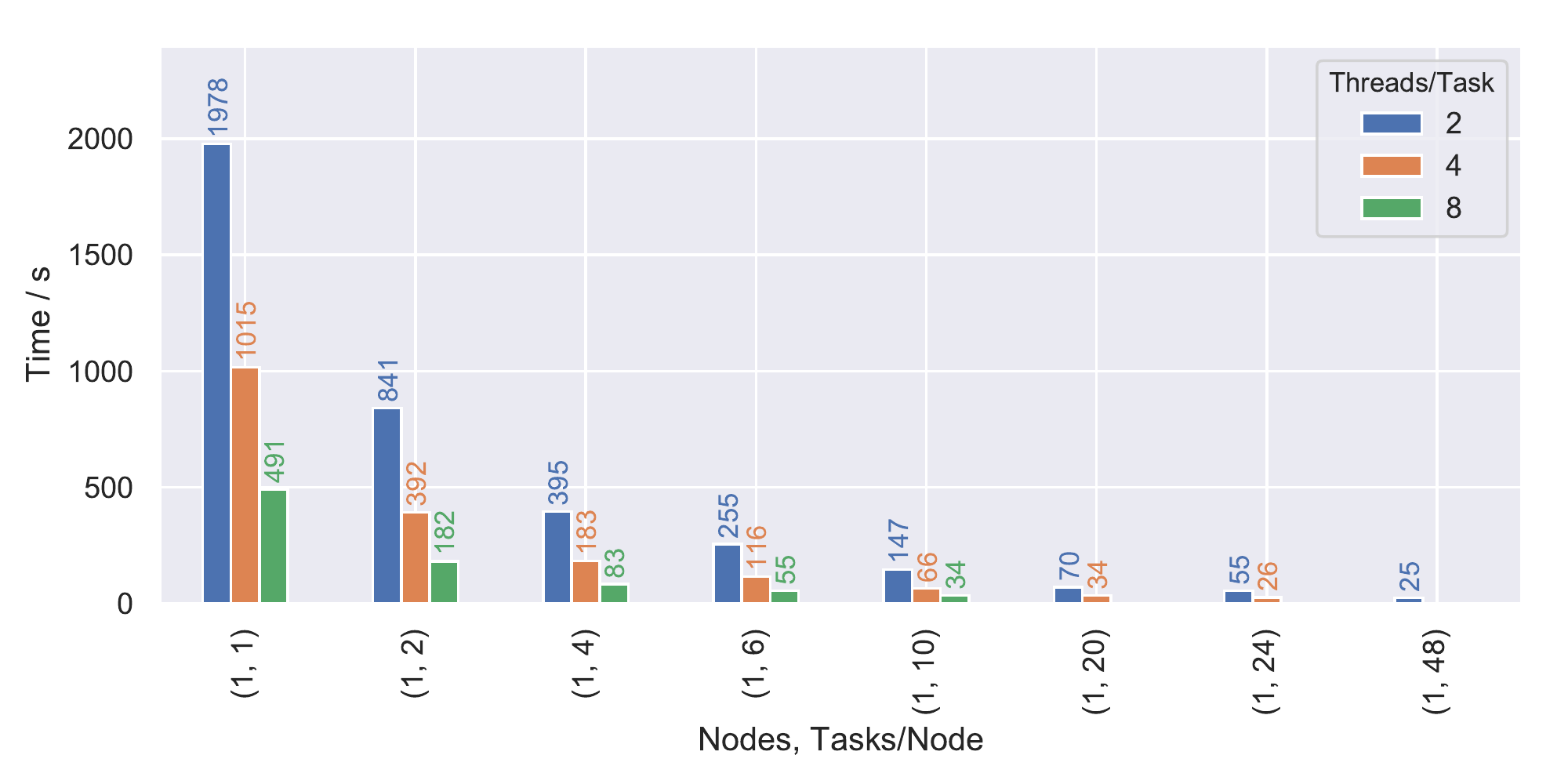}
  \caption{NEST benchmark -- JUWELS: Same as in Fig.~\ref{fig:nest:singlenode:juron} but for JUWELS.}
  \label{fig:nest:singlenode:juwels}
\end{figure}

In Fig.~\ref{fig:nest:singlenode:juron} and \ref{fig:nest:singlenode:juwels} we show how
simulation time scales on a single node as a function of the number of VPs.
The number of VPs is equal to the number of threads, which is
the product of the number of nodes, tasks per node, and threads per task.
The results for multi-node scaling are shown in Table~\ref{tab:nest:multinode}.

\begin{table}[t]
  \setlength{\tabcolsep}{7pt}
  \caption{NEST benchmark: Build and simulation time for 1, 2 and 4 nodes
           and optimal number of tasks and threads per node.}
  \label{tab:nest:multinode}
  \begin{center}
  \begin{tabular}{c@{\hspace{2em}}cc@{\hspace{2em}}cc}
  \toprule
           & \multicolumn{2}{c}{JURON} & \multicolumn{2}{c}{JUWELS}    \\
    Nodes  & Build / \si{\second} & Simulation / \si{\second} & Build / \si{\second} & Simulation / \si{\second}   \\
  \midrule
  1               & 2.58       & 44.50           & 1.25       & 25.15               \\
  2               & 2.34       & 32.39           & 0.81       & 15.15	               \\
  4               & 1.39       & 18.80           & 0.63       & 5.88               \\
  \bottomrule
  \end{tabular}
  \end{center}
\end{table}

NEST can efficiently exploit node- and thread-level parallelism and therefore exhibits
a good scaling behaviour on both architectures.
On POWER8 processors the application to some extent benefits from using up to 8 hardware
threads per core.
This trend can not compensate for the larger number of cores available on the Skylake processor.
Since NEST is not capable of exploiting SIMD parallelism, the wider SIMD units of the Xeon
processors do not add benefits for this application.

\begin{figure}[ht]
  \centering
  \includegraphics[width=\textwidth]{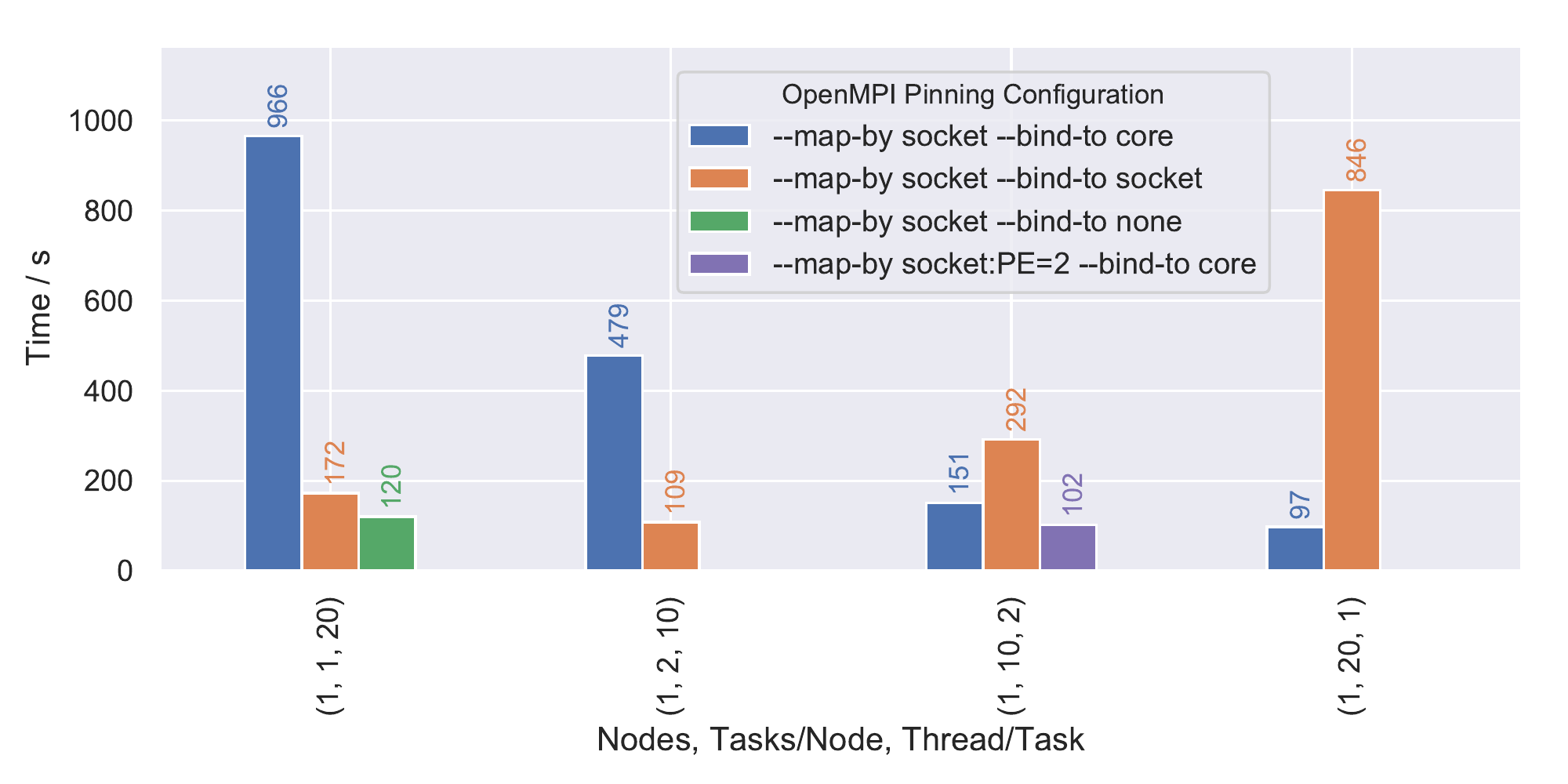}
  \caption{NEST benchmark -- JURON: Different binding strategies for different distributions of VPs.}
  \label{fig:nest:pinning:juron}
\end{figure}

For NEST it is of special importance how the VPs are distributed to the available hardware resources. NEST uses OpenMP for shared-memory parallelism (\emph{Threads/Task}) and MPI for task-based parallelism (\emph{Tasks/Node}), potentially across node borders. The employed strategy to distribute OpenMP threads uses the \verb|OMP_PLACES| environment variable to distribute along physical cores (\texttt{cores}); this variable might possibly interact with MPI binding options. To fix each OpenMP thread to a certain place, the affinity variable \verb|OMP_PROC_BIND| is set to \texttt{TRUE}. Of more importance for the two-socket system JURON is the employed distribution of MPI tasks. In Fig.~\ref{fig:nest:singlenode:juron}, two binding schemes of OpenMPI were used. The hatched bars (usually for lower number of VPs) bind tasks to cores (\verb|--bind-to core|) whereas the solid bars bind tasks to sockets (\verb|--bind-to socket|); in both cases, a mapping along sockets is chosen (\verb|--map-by socket|). Setting the \emph{wrong} binding can entail serious performance penalties (see also the white-outlined bars in the background of Fig.~\ref{fig:nest:singlenode:juron}). Fig.~\ref{fig:nest:pinning:juron} compares the two binding strategies for four selected distributions of 20 VPs along tasks and threads. It can clearly be seen that for few tasks and many threads in the left of the figure, \verb|--bind-to socket| is the more beneficial binding option. For many tasks with each few threads (right of the figure), \verb|--bind-to core| is the more sensible choice. While these results might be expected, they might not be the choice of distribution for the employed MPI. Fig.~\ref{fig:nest:pinning:juron} also shows two further optimized binding configurations: For the configuration of 1 Task and 20 Threads (the very left), disabling binding can improve performance further (\verb|--bind-to none|) as now both sockets can be used by the 20 threads. Another way to bind can be seen for the case of 10 Tasks and 2 Threads per Task; by binding to cores but also mapping to the sockets such that exactly 2 cores (\verb|PE=2|) are bound to each task, the MPI runtime has all information about tasks and threads to create a performance-beneficial task distribution (each task has two associated OpenMP places).\\
The mapping and binding options for JUWELS are more limited as the combination of job scheduler and MPI runtime currently don't offer similar high-level functionality as in the case for JURON. For JUWELS, the shown values are measured by using a pinning mask manually created by the tool \verb|hwloc|, distributing tasks across the node architecture. In general, the default pinning of JUWELS is much better than the default pinning of JURON.

\begin{figure}[ht]
  \centering

  \includegraphics[width=0.8\textwidth]{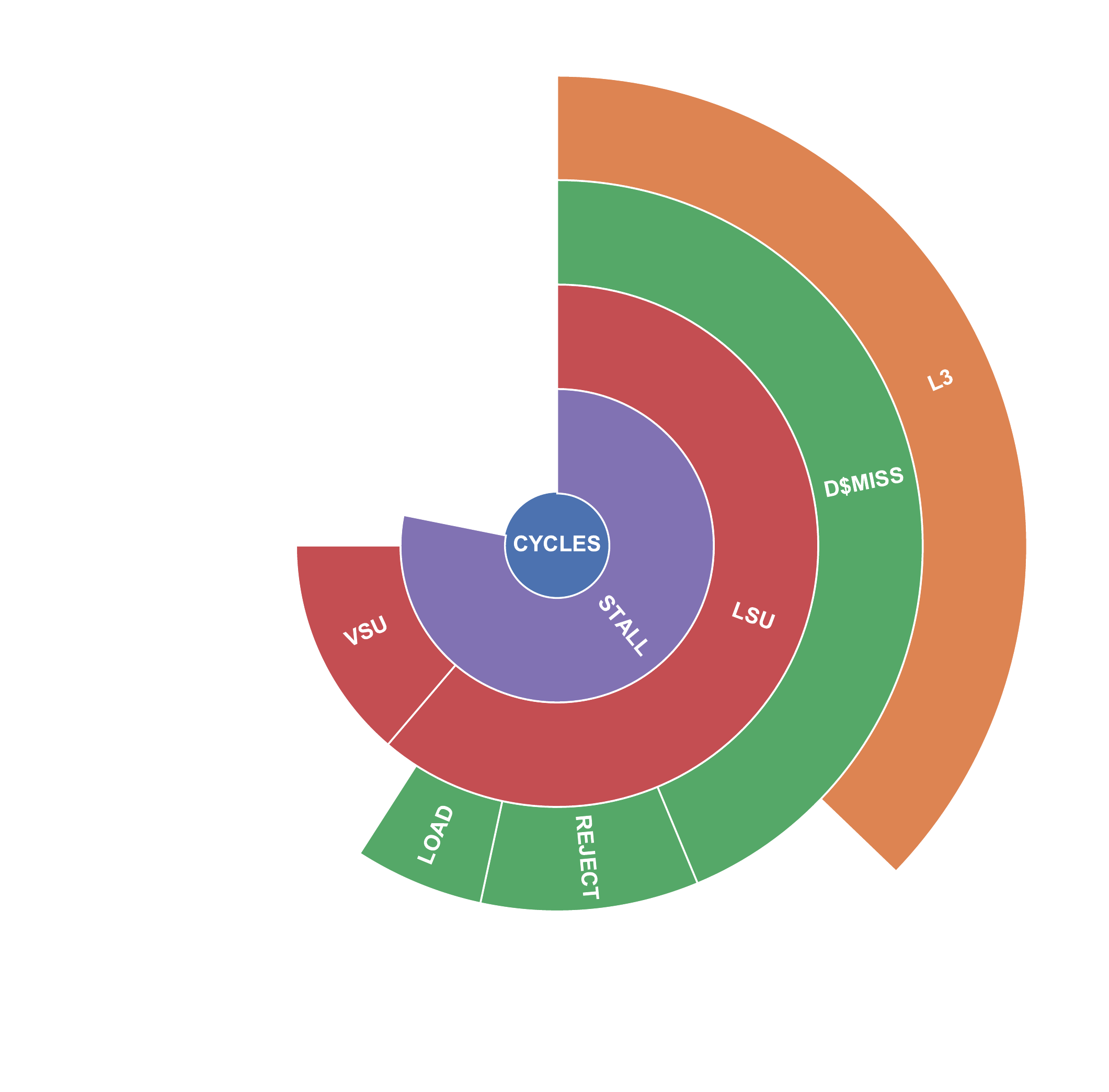}
  \caption{NEST benchmark -- JURON: Performance counters.}
  \label{fig:nest:counter:juron}
\end{figure}

One of the main performance limiters for NEST on JURON are stalled processor cycles, as shown in Fig.~\ref{fig:nest:counter:juron}. The figure shows measurements of a selection of hardware performance counters using the \verb|perf| utility. Most of the stalls can be attributed to misses of the data cache, more specifically to cases were data was not in the L3 level cache.

\subsection{Arbor}

Arbor is a rather new simulator, which has been designed from scratch with the goal of supporting
different HPC architectures in a performance portable manner.
We use version 0.1 of Arbor\cite{arborv1}.\footnote{\url{https://github.com/arbor-sim/arbor.git}}
The different simulation phases are similar as for NEST, but given that simulation of
multi-compartment models of neurons are much more expensive, the benchmark focuses exclusively
on the simulation phase.
Simulation proceeds in a lock-step manner by first updating all cells half a time step and then overlapping the exchange of spikes with the further half time steps.
Arbor allows to group cells in a flexible manner depending on the target architecture.
For instance, large cell groups are used for GPUs that require a very high level of parallelism.
For this paper the GNU C++ compiler version 6.3.0 and 8.2.0 have been used on JURON and JUWELS, respectively, when running the CPU version of the benchmark, and 6.3.0 and 7.3.0, respectively, when running the GPU version of the benchmark. For the GPU benchmark CUDA 9.2.148 was used on JURON and CUDA 9.2.88 was used on JUWELS.
We perform simulations involving \SI{1000}{cells} covering a biological time of 
\SI{1}{\second} (GPU version) or \SI{0.1}{\second} (CPU version).

\begin{figure}[th]
    \centering
    \includegraphics[width=\textwidth]{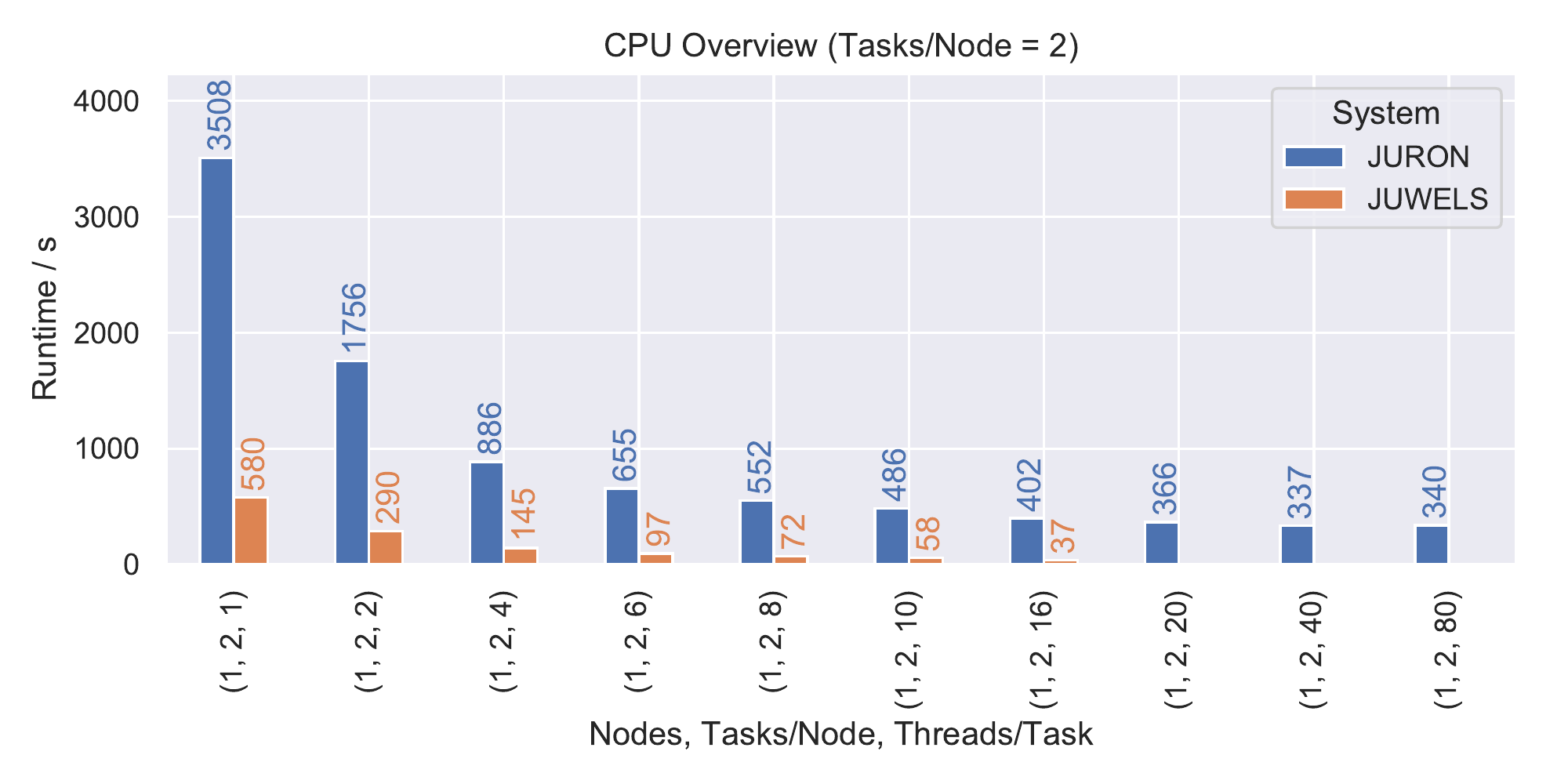}
    \caption{Arbor benchmark: Time needed to simulate \SI{100}{\milli\second} biological time on a single 
    CPU-only node.}
    \label{fig:arbor:cpu}
\end{figure}

In Fig.~\ref{fig:arbor:cpu} we show how performance scales on a single CPU-only node
using 2 MPI tasks and a variable number of threads per MPI task.
On POWER8 little benefit is observed from using multiple hardware threads per physical CPU core.
The application is significantly faster on the Xeon nodes as it can efficiently exploit the
wide SIMD units.
The observed performance ratio roughly matches the ratio of throughput in floating-point
operations (see Table~\ref{tab:hwpar}).
The higher clock frequency of the POWER8 processor does not seem to help improving performance
significantly.

\begin{figure}[th]
    \centering
    \includegraphics[width=\textwidth]{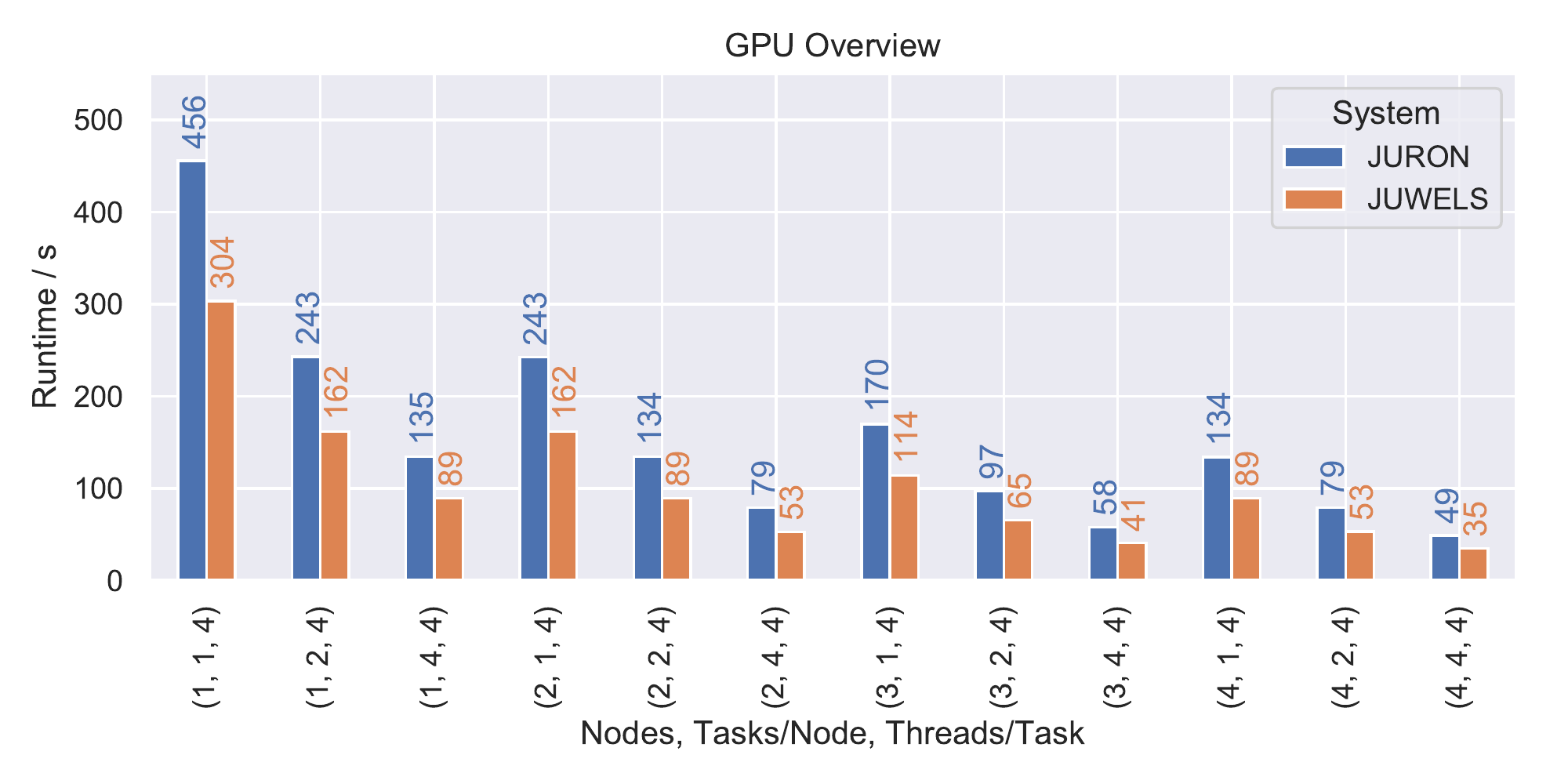}
    \caption{Arbor benchmark: Time needed to simulate \SI{1000}{\milli\second} biological time on
    one or more GPU-accelerated nodes.}
    \label{fig:arbor:gpu}
\end{figure}

Next we compare performance using Arbor on GPU-accelerated nodes.
The results are shown in Fig.~\ref{fig:arbor:gpu}.
To highlight differences in scaling among different numbers of GPUs, the simulation time has 
been increased ten times compared to the CPU-only benchmark.
Performance on JUWELS and JURON now is similar when taking into account that the V100 GPUs used
in JUWELS have an about \SI{40}{\percent} higher throughput of double-precision floating-point operations
compared to the P100 GPUs used in JURON.

\subsection{TVB-HPC}

The HPC version of TVB, which uses Numba for code generation,
is still in development and we therefore use a pre-release version.
The benchmark uses a simple mesoscopic model based on the Kuramoto model \cite{10.1007/BFb0013365},
which is used for the study of neuronal oscillations and synchronization.
\num{1600} time steps are simulated.
We use Python 3.6.1 (3.6.6) as well as 
version 0.39.0 (0.40.1) of Numba and
version 1.14.2 (1.15.2) of Numpy on JURON (JUWELS).

\begin{figure}[th]
    \centering
    \includegraphics[width=\textwidth]{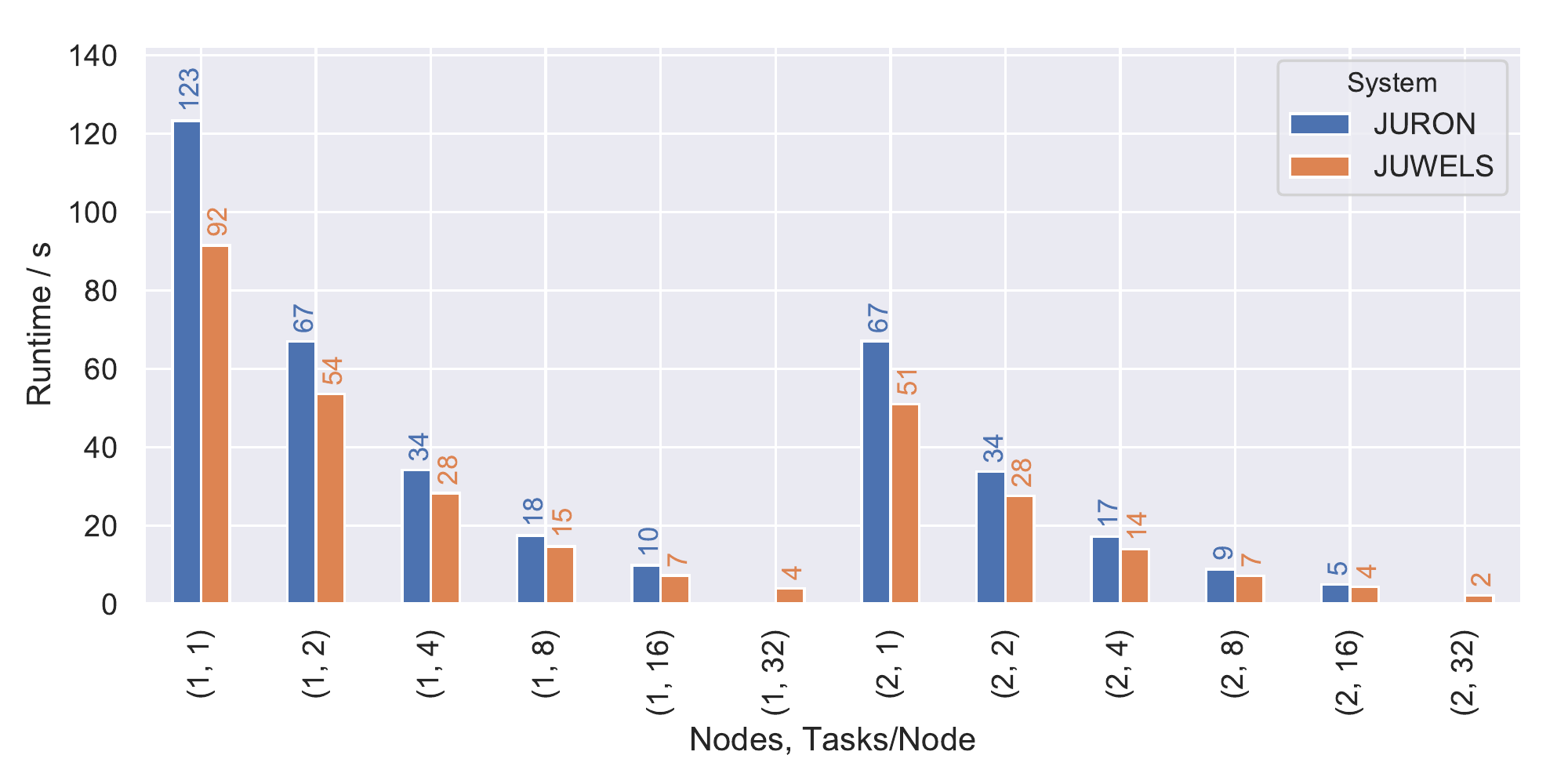}
    \caption{TVB-HPC benchmark: Scaling of the benchmark on up to 2 CPU-only nodes as a function
             of the number of MPI tasks.}
    \label{fig:tvb-hpc:scaling}
\end{figure}

In Fig.~\ref{fig:tvb-hpc:scaling} we show the scaling of the benchmark on up to two nodes.
The observed scaling behaviour is similar on both architectures with a parallel efficiency
of about \SI{80}{\percent} when using 16 MPI tasks on a single node.
On a single node the benchmark execution time on JURON is consistently about \SI{35}{\percent} slower
compared to JUWELS when using the same number of MPI tasks.
The difference drops to about \SI{13}{\percent} when using two nodes.
This version of TVB-HPC is not able to exploit SIMD parallelism.
This observation plus the limited scalability results in performance differences between
JURON and JUWELS that are relatively small.

\subsection{Elephant ASSET}

The benchmark is based on version 1.0 of ASSET%
\footnote{\url{https://github.com/INM-6/pcp_use_cases.git}}
as well as Python libraries
neo (version 0.6.1), sklearn (version 0.19.1), and elephant (version 0.5.0), respectively.
Most of the time is spent computing a set of survival functions, which requires computing
of statistical distributions based on the input data.
This makes speed of memory access in general a performance limiting factor,
with efficiency of handling of Python arrays being an implementation specific aspect.
Parallelisation is realised by distributing the input data in an approximately fair manner
to all available MPI tasks.
While MPI parallelisation is done explicitly within the application, exploitation
of any additional thread-level parallelism is left to Python.

\begin{figure}[th]
    \centering
    \includegraphics[width=\textwidth]{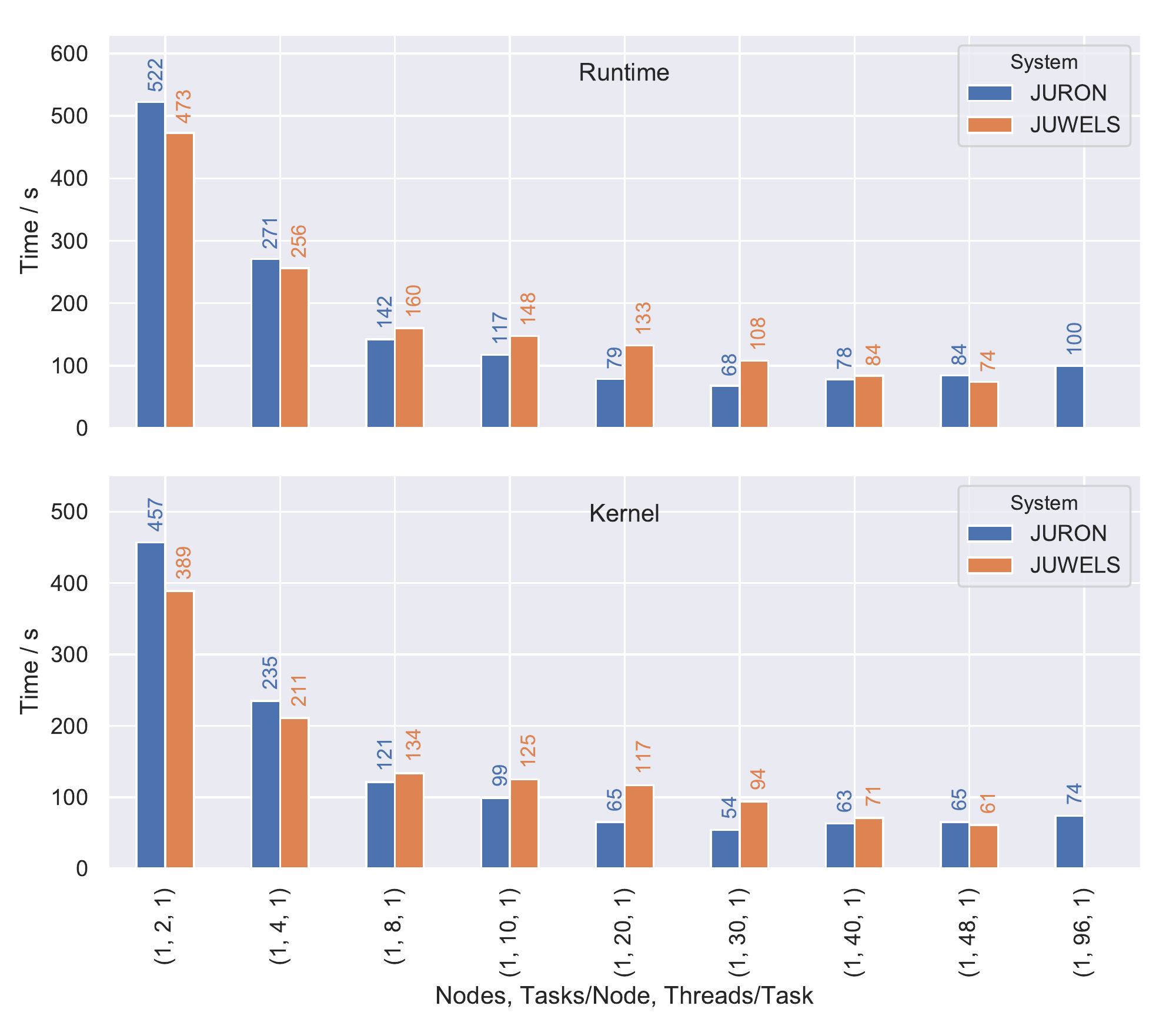}
    \caption{Elephant-ASSET benchmark: Benchmark (top) and kernel (bottom) execution times as a function of tasks on one node for JURON and JUWELS.}
    \label{fig:asset:scaling}
\end{figure}

In Figure~\ref{fig:asset:scaling} we show benchmark run-time as well as the execution time
of the main kernel, which computes the joint survival function,
as a function of the number of tasks.
While execution time on JUWELS is first lower than on JURON, the scaling behaviour on JUWELS is slightly worse, resulting in JURON having the lowest achieved benchmark time with a sufficient number of tasks (30).

\subsection{Neuroimaging Deep Learning}

The benchmark is a mini-application version of the real code.
It is extracted such that input data is loaded first to main memory to separate performance
impacts related to the capabilities of the storage system, in which the input data resides.%
\footnote{I/O performance for this application is crucial and has been analysed in 
\cite{DBLP:conf/pdp/OdenSSDP19}.}
For the results shown below TensorFlow version 1.4.1 (1.8.0) and Horovod version 0.14.1 (both) was used on JURON (JUWELS).

\begin{figure}[th]
    \centering
    \includegraphics[width=\textwidth]{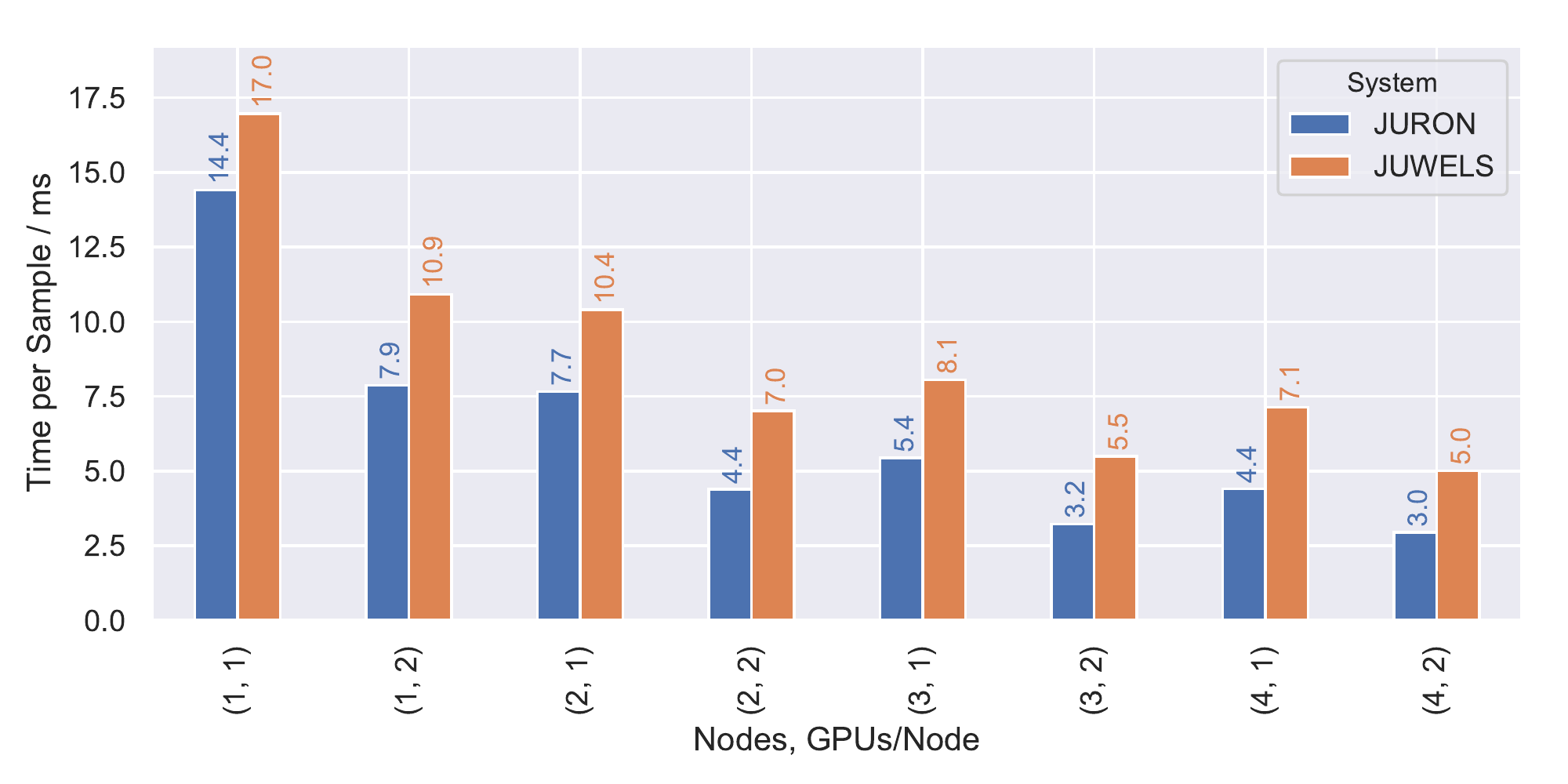}
    \caption{Neuroimaging Deep Learning benchmark: Time per sample as a function of GPUs and tasks for JURON and JUWELS.}
    \label{fig:neuroimg:scaling}
\end{figure}

A selection of benchmark results are listed in Figure~\ref{fig:neuroimg:scaling}.
The performance is measured in terms of time needed to process a single image.
On both of considered architectures the intra-node scaling as well as the inter-node
scaling behaviour is fair. The performance on JUWELS suffers efficiency losses of (max.) \SI{30}{\percent} and \SI{23}{\percent} for intra- and inter-node scaling, respectively; for JURON it is \SI{10}{\percent} and \SI{6}{\percent}.
Absolute performance and scaling performance is significantly better on JURON despite an older generation of 
GPUs being used.
This may be an indication of better data transport capabilities having an important
performance impact in this application.

\section{Summary and Conclusions}\label{sec:summary}

In this paper we presented selected performance results obtained for the ``ICEI Application
Benchmark Suite'' using a slightly older system based on IBM POWER8 processors plus NVIDIA P100
GPUs (\emph{JURON}) as well as a more recent system based on Intel Skylake processors plus NVIDIA V100 GPUs (\emph{JUWELS}).

The example of Arbor indicates
that for compute-limited applications, which can exploit wide SIMD pipelines without using GPUs
as compute accelerators, the performance of CPU-only JUWELS nodes exceed those of JURON
significantly. For processors based on the POWER architecture being competitive, higher
throughput of floating-point operations would be desirable.

For applications, which cannot exploit SIMD parallelism, performance on JURON and JUWELS was
found to be similar in case of TVB-HPC and ASSET.
In case of NEST, which compared to the other applications is able to scale slightly
better by making use of the larger number of available cores, JUWELS was found to perform better.

Finally, for the machine learning application the POWER-based system was found to be better.

The results provided in this paper give an overview over the performance trend for different
applications from a benchmark suite that aims for reflecting the needs of the relatively diverse
community of brain research.
We plan for further efforts to analyse the causes for the different performance behaviour.
This will help to guide designing future e-infrastructures optimised for this community.
We believe that the increased choice of architectures and technologies, which can be used for 
this purpose, is helpful.

\section*{Acknowledgements}

We would like to thank the many people that have contributed to the creation of the ICEI Benchmark
Suite, which was used for this paper.
This includes in particular the following persons:
Ben Cumming (CSCS, Switzerland),
Sandra Diaz (JSC, Germany),
Pramod Kumbhar (EPFL, Switzerland),
Lena Oden (JSC and FU Hagen, Germany),
Alexander Peyser (JSC, Germany),
Hans Ekkehard Plesser (NMBU, Norway),
Alper Yegenoglu (FZJ, Germany),
and all the collaborators of the respective community codes used.
Funding for the work is received from the European Union’s Horizon 2020 research and innovation programme under grant agreement No.~785907 (HBP SGA2) and No.~800858 (ICEI).

\bibliographystyle{splncs03}
\bibliography{iwoph19-icei}

\end{document}